Two-dimensional extremely short optical pulses in silicene with random electric fields

N.N. Konobeeva[1], M.B. Belonenko[1,2]

[1] Volgograd State University, 400062, Volgograd, Russia

[2] Volgograd Institute of Business, Volgograd, Russia

E-mail: yana_nn@inbox.ru

We investigate an influence of the random external electric field on the 2D extremely short optical pulses propagation in the silicene. The random electric field is perpendicular to the silicene plane. An effective equation has the form of a wave equation with saturating nonlinearity. We analyze the dependences of the electromagnetic field intensity on the parameters of a problem.

### Introduction

In recent years, the nonlinear propagation of light in graphene structures has been intensively studied. These structures have unique electrical properties for modern micro- and nanoelectronics [1]. One of these materials is silicene—a monolayer of silicon atoms with a hexagonal lattice [2-3], which is characterized by a stronger spin–orbit interaction as compared to that in graphene. One of the most interesting predictions for silicene is the appearance of the band gap, which can lead to the appearance of the transition between the itinerant and the topological insulator. In addition, it is important that silicon is still the main element in modern electronics.

In recent work, we studied the issue of one-dimensional propagation of extremely short electromagnetic pulses in silicene waveguides [4]. It has been shown, that at the certain moment there is an inversion signal. The amplitude of the inverted signal is almost twice as large as that of the initial signal. Thus, one can speak of an amplification of ultimately short optical pulses with a sharp change their form. At the same time there are many questions related to the study of multi-dimensional (2D and 3D) pulses in silicene.

It is possible to control the width of the band gap with an electric field in the perpendicular direction to the plane, because of atoms in silicene do not lie exactly in one plane, and are located above and below it. In this case, the potential difference between the sublattices of the gap depends directly on applied dc electric field. It is obvious that, an important role has a random electric field applied as described above. Random fields can arise from both the charged impurities between the silicene, planes so, and for example, have an external nature. In this case, the electric field pulse will propagate in a medium with random bandgap and earlier findings should be clarified.

**Statement of a problem**

Taking into account these circumstances, it is important to study the interaction of extremely short optical pulses with silicone in situation with random electric fields, which is expected to reveal new effects with a broad range of practical applications.

In a long-wavelength approximation, a Hamiltonian for silicene can be written in the following form [8, 9]:

$$H=v(\xi k_x \sigma_x + k_y \sigma_y) - 0.5\xi\Delta_{SO}\tau_z\sigma_z + 0.5\Delta_z\sigma_z \qquad (1)$$

where $\xi - \pm$ is the sign of the valley for two Dirac dots, $v$ is the Dirac electron velocity, $\boldsymbol{p}$=(k$_x$, k$_y$) – is the quasi-momentum of electrons, $\Delta_{SO}$ is the spin–orbit gap width in silicene, $\Delta_z$ is the potential on a lattice site, причем $\Delta_z$=$E_z$d, $E_z$ is the constant random electric field, d is the distance between two sublattice planes, $\sigma_i$, $\tau_i$ are the Pauli matrices. Field $E_z$ is random with a probability density:

$w(E_z) = \dfrac{1}{\sigma\sqrt{2\pi}}\exp(-\dfrac{(E_z - E_{z0})^2}{2\sigma^2})$. Here $E_{z0}$ is the average value of the random field, $\sigma$ is the standard deviation.

Next, let us consider the problem for a single layer of silicene, where the field $E_z$ has a certain value, and then we make an averaging, taking into account that extremely short pulse has a sufficient size of spatial localization for averaging on random fields. This imposes a limit on the average frequency of extremely short pulses and its spectrum.

The eigen values of matrix form for the Hamiltonian is:

$$\varepsilon_{\sigma\xi} = \pm\sqrt{v^2k^2 + \frac{1}{4}\left(\Delta_z - \sigma\xi\Delta_{SO}\right)^2} \qquad (2)$$

where σ is the electron spin (spin «up» и «down»).

The Maxwell equations with taking into account: $\boldsymbol{E} = -\dfrac{\partial \boldsymbol{A}}{c\,\partial t}$ and with a replacement momentum on generalized momentum: $p \rightarrow p - eA/c$ (e is the electron charge), in two-dimensional case can be written as [10]:

$$\frac{\partial^2 \boldsymbol{A}}{\partial x^2} + \frac{\partial^2 \boldsymbol{A}}{\partial z^2} - \frac{1}{c^2}\frac{\partial^2 \boldsymbol{A}}{\partial t^2} + \frac{4\pi}{c}\boldsymbol{j} = 0 \qquad (4)$$

where vector-potential $\boldsymbol{A}$ has the form $\boldsymbol{A}$=(0, A(x,z,t),0), and current density $\boldsymbol{j} = (0, j, 0)$.

Let us write the standard expression for the current density for one silicene layer $j_s$:

$$j_s = e\sum_p v_y(p - \frac{e}{c}A(x,z,t))\left\langle a_p^+ a_p \right\rangle, \qquad (5)$$

where $v(p)=\partial\varepsilon(p)/\partial p$, and the angle brackets denote the averaging with the nonequilibrium density matrix $\rho(t)$: $\langle B\rangle = Sp(B(0)\rho(t))$, $a_p^+, a_p$ are the creation and annihilation operators of quasimomentum $p$.

Let us restrict consideration to the case of low temperatures, whereby only a small region near the Fermi level in the space of momenta contributes to the sum in (5). Accordingly, formula (6) can be rewritten as:

$$j_s = e\int\limits_{-\Lambda}^{\Lambda}\int\limits_{-\Lambda}^{\Lambda} dp_z dp_y v_y (p - \frac{e}{c}A(x,z,t)) \qquad (6)$$

The integration domain is determined from the condition of conservation of the number of particles:

$$\int\limits_{-\Lambda}^{\Lambda}\int\limits_{-\Lambda}^{\Lambda} dq_z dq_y = \iint\limits_{ZB} dq_z dq_y \langle a_{qz,qy}^+ a_{qzqy}\rangle$$

where the integral on the right-hand side is taken over the first Brillouin zone,

The propagation of an ultimately short pulse is described by the following equation:

$$\frac{\partial^2 A}{\partial x^2} + \frac{\partial^2 A}{\partial z^2} - \frac{1}{c^2}\frac{\partial^2 A}{\partial t^2} + \frac{4\pi}{c}\Phi(A) = 0\,, \qquad (7)$$

where $\Phi(A)$ is determined by integral (6) and by averaging of the current on random fields $E_z$:

$$\Phi(A) = \int\limits_{-\infty}^{\infty} j_s(A)w(E_z)dE_z$$

**Main results**

Equation (7) has been numerically solved [11]. The initial conditions in 2D case were taken in the form of a Gaussian pulse profile:

$$A(z,0) = Q\cdot exp\left(-z^2/\gamma_z^2\right)exp\left(-x^2/\gamma_x^2\right),$$
$$\frac{dA(z,0)}{dt} = \frac{2z\,v_z}{\gamma_z^2}Q exp\left(-z^2/\gamma_z^2\right)exp\left(-x^2/\gamma_x^2\right) \qquad (9)$$

here Q is the pulse amplitude, $v_{z,x}$ is the initial velocity along z and x axiz, $\gamma_{z,x}$ is the pulse width. Time has been chosen as an evolution coordinate.

The values of energy parameters are expressed in units of $\Delta$.

The evolution of a two-dimensional electromagnetic field propagating in sample is shown in Fig. 1.

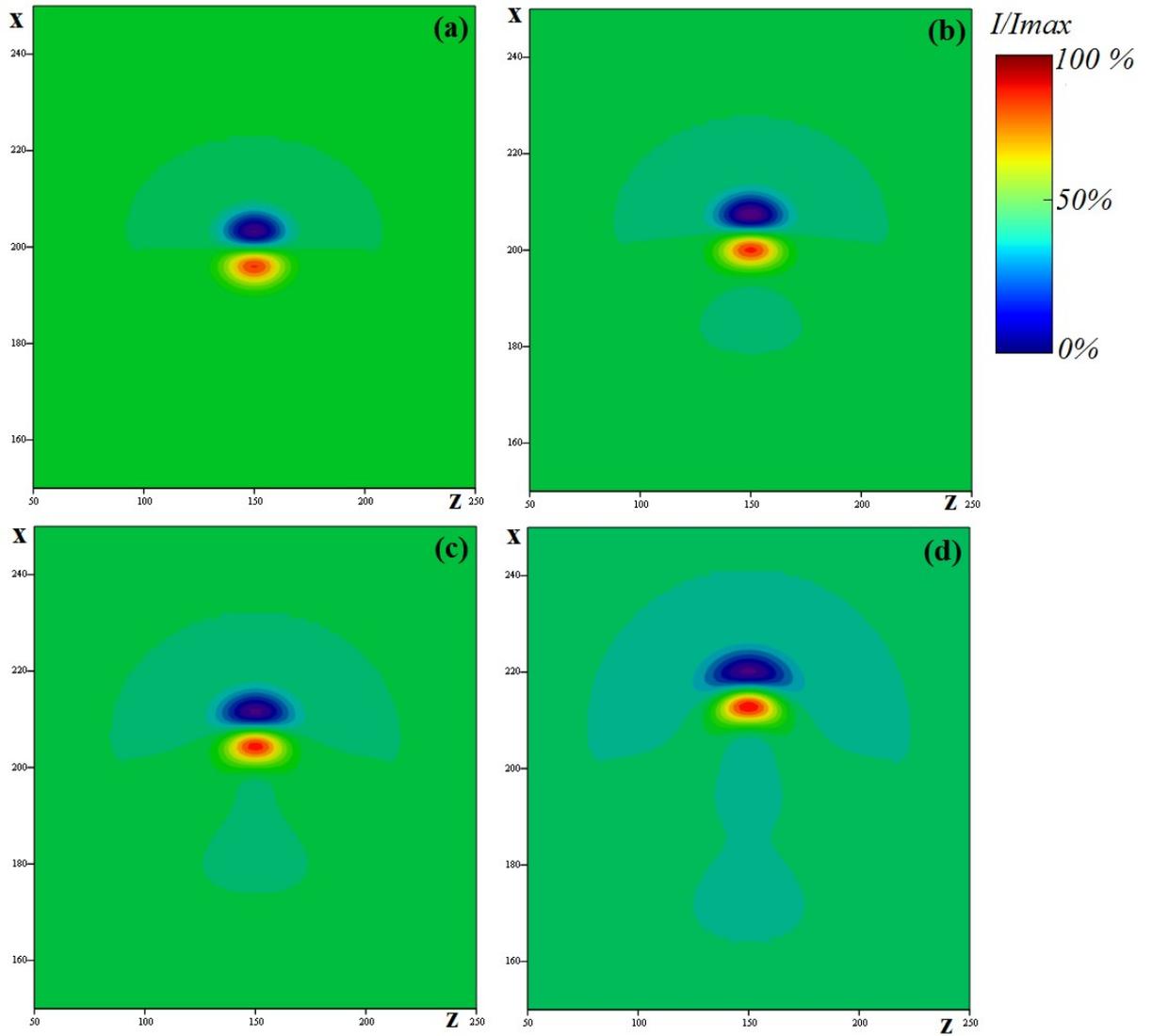

Fig.1. The intensity of a two-dimensional electromagnetic pulse $I(x,z,t)=E^2(x,z,t)$ at different instances of time: a) initial pulse form; b) t=0.5·10^{-13} s; c) t=1.0·10^{-13} s; d) t=5.0·10^{-13} s.

Form of the main pulse form almost does not change. It can be seen only a slight distortion of the wave front, herewith an amplitude is preserved. It should be noted, that "tail" after the main pulse increases over time. This fact may be associated with the excitation of the nonlinear wave pulses.

Comparison of the case of an external alternating electric field $E_z$ and without it is shown in Fig. 2.

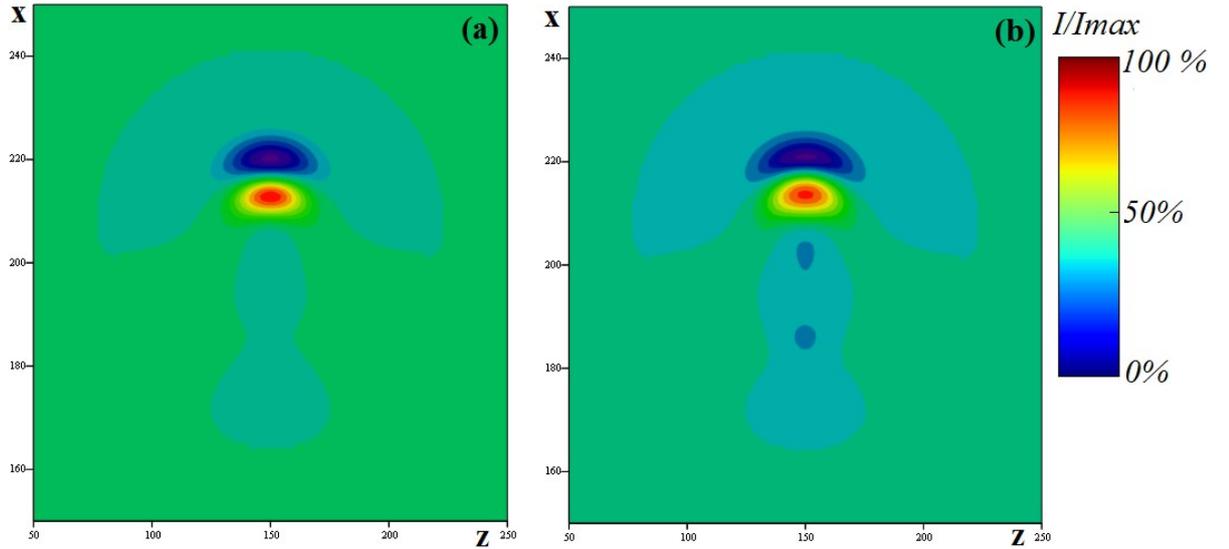

Fig.2. The intensity of a two-dimensional electromagnetic pulse $I(x,z,t)=E^2(x,z,t)$ at the moment t=2.0·10$^{-13}$ s: a) with averaging of field $E_z$; b) without averaging.

As can be seen in Fig.2, takin into account averaging of the external electric field has a great influence on the pulse propagation, which manifests in decreasing of a "tail" behind the main pulse and in more pulse amplitude than in the case (b).

Thus, the introduction of an electric field averaging provides the balance between dispersion and nonlinearity of the system, which leads to the amplitude preservation and stabilization of the pulse.

Our calculation results show that in contrast to the case without averaging of the external electric field, the influence of its dispersion is weak.

**Conclusion**

In conclusion, we formulate the main outcomes of the work done.

1. Averaging of the electric field provides a balance between dispersive effects and nonlinear ones for this system. Thus, we can speak about stable pulse propagation in silicene.

2. Extremely short optical pulses cause the appearance of a "tail" that can be associated with the excitation of nonlinear waves.

3. Extremely short pulse propagation in silicene based on the external alternating field $E_z$ averaging is more stable in comparison with the case without averaging.

**Acknowledgments.**